\documentclass[aps,prb,reprint,superscriptaddress,floatfix,showpacs]{revtex4-1}
\usepackage{color}
\usepackage{amssymb}
\usepackage{amsmath}
\usepackage{textcomp}
\usepackage{graphicx}
\usepackage{float}
\usepackage{natbib}
\usepackage{url}
\usepackage{array}
\usepackage{multirow}
\usepackage[bookmarksnumbered,pdfpagelabels=true,plainpages=false,colorlinks=true,linkcolor=blue,citecolor=blue,urlcolor=blue]{hyperref}

\begin{document}
\title{The role of the internal demagnetizing field for the dynamics of a magnonic crystal} %Title of paper
%
%\title{Extracting nanoscale properties of magnonic crystals}
%
\author{M.\ Langer}
\affiliation{Helmholtz-Zentrum Dresden -- Rossendorf, Institute of Ion Beam Physics and Materials Research, Bautzner Landstr.\ 400, 01328 Dresden, Germany}
\affiliation{Institute for Physics of Solids, Technische Universit\"at Dresden, Zellescher Weg 16, 01069 Dresden, Germany}
\author{F.\ R\"oder}
\affiliation{Helmholtz-Zentrum Dresden -- Rossendorf, Institute of Ion Beam Physics and Materials Research, Bautzner Landstr.\ 400, 01328 Dresden, Germany}
\affiliation{Triebenberg Laboratory Institute of Structure Physics, Technische Universit\"at Dresden, 01062 Dresden, Germany}
\affiliation{{CEMES-CNRS and Universit\'{e} de Toulouse, 29 Rue Jeanne Marvig, F-31055 Toulouse, France}}
\author{R.\ A.\ Gallardo}
\affiliation{Departamento de F\'isica, Universidad T\'ecnica Federico Santa Mar\'ia, Avenida Espa$\tilde{n}$a 1680,
2390123 Valpara\'iso, Chile}
\author{T.\ Schneider}
\affiliation{Helmholtz-Zentrum Dresden -- Rossendorf, Institute of Ion Beam Physics and Materials Research, Bautzner Landstr.\ 400, 01328 Dresden, Germany}
\affiliation{Department of Physics, Technische Universit\"at Chemnitz, Reichenhainer Str.\ 70, 09126 Chemnitz, Germany}
\author{S.\ Stienen}
\affiliation{Helmholtz-Zentrum Dresden -- Rossendorf, Institute of Ion Beam Physics and Materials Research, Bautzner Landstr.\ 400, 01328 Dresden, Germany}
\author{C.\ Gatel}
\affiliation{{CEMES-CNRS and Universit\'{e} de Toulouse, 29 Rue Jeanne Marvig, F-31055 Toulouse, France}}
\author{R.\ H\"ubner}
\affiliation{Helmholtz-Zentrum Dresden -- Rossendorf, Institute of Ion Beam Physics and Materials Research, Bautzner Landstr.\ 400, 01328 Dresden, Germany}
%
%\author{A.\ Rold\'an-Molina}
%\affiliation{Instituto de F\'isica, Pontificia Universidad Cat\'olica de Valpara\'iso, Avenida Brasil 2950,
%2390123 Valpara\'iso, Chile}
%
\author{L.\ Bischoff}
\affiliation{Helmholtz-Zentrum Dresden -- Rossendorf, Institute of Ion Beam Physics and Materials Research, Bautzner Landstr.\ 400, 01328 Dresden, Germany}
\author{K.\ Lenz}
\affiliation{Helmholtz-Zentrum Dresden -- Rossendorf, Institute of Ion Beam Physics and Materials Research, Bautzner Landstr.\ 400, 01328 Dresden, Germany}
\author{J.\ Lindner}
\affiliation{Helmholtz-Zentrum Dresden -- Rossendorf, Institute of Ion Beam Physics and Materials Research, Bautzner Landstr.\ 400, 01328 Dresden, Germany}
\author{P.\ Landeros}
\affiliation{Departamento de F\'isica, Universidad T\'ecnica Federico Santa Mar\'ia, Avenida Espa$\tilde{n}$a 1680,
2390123 Valpara\'iso, Chile}
\author{J.\ Fassbender}
\affiliation{Helmholtz-Zentrum Dresden -- Rossendorf, Institute of Ion Beam Physics and Materials Research, Bautzner Landstr.\ 400, 01328 Dresden, Germany}
\affiliation{Institute for Physics of Solids, Technische Universit\"at Dresden, Zellescher Weg 16, 01069 Dresden, Germany}
\date{\today}
\begin{abstract}
Magnonic crystals with locally alternating properties and specific periodicities exhibit interesting effects, such as a multitude of different spin-wave states and large band gaps. This work aims for demonstrating and understanding the key role of local demagnetizing fields in such systems. To achieve this, hybrid structures are investigated consisting of a continuous thin film with a stripe modulation on top favorable due to the adjustability of the magnonic effects with the modulation size. For a direct access to the spin dynamics, a magnonic crystal was reconstructed from `bottom-up', i.e., the structural shape as well as the internal field landscape of the structure were experimentally obtained on the nanoscale using electron holography. Subsequently, both properties were utilized to perform dynamic response calculations.
The simulations yield the frequency-field dependence as well as the angular dependence of spin waves in a magnonic crystal and reveal the governing role of the internal field landscape around the backward-volume geometry. The complex angle-dependent spin-wave behavior is described for a 360$^\circ$ in-plane rotation of an external field by connecting the internal field landscape with the individual spin-wave localization.
\end{abstract}
\newcommand{\todo}[1]{\textbf{\textcolor{red}{\emph{TODO:\newline}#1}}}
\newcommand{\new}[1]{\textcolor{blue}{#1}}
\maketitle 
\section{Introduction}
\label{Int}
Magnetic meta-materials, especially magnonic crystals (MCs),\cite{Kruglyak2010,Gubbiotti2010,Lenk2011,Krawczyk2014,Chumak2015} experience a growing scientific attention due to many promising applications for future devices in information technology. The root of this development lies in the unique properties of MCs,\cite{Tacchi2012,Krawczyk2013,Montoncello2013} such the large band gaps and a multitude of adjustable magnon bands.\cite{Kostylev2008,Lee2009,Wang2009,Wang2010,Ma2012,Krawczyk2013,Kumar2014} Both can be engineered or even tuned by modifying their structural or magnetic properties.\cite{Chumak2010,Vogel2015} In addition, MCs, in particular one-dimensional systems, possess the possibility of reprogramming the magnonic properties by a switching between different states in the magnetic hysteresis.\cite{Topp2010,Tacchi2010,Topp2011,Ding2011,Lin2012,Di2015}
In previous studies, it was already shown that MCs can be used as grating couplers,\cite{Yu2013} for magnonic logic,\cite{Kostylev2005,Lee2008,Khitun2010,Vogt2014} filter\cite{Kim2009} and sensor\cite{Inoue2011} applications, and moreover, as a tool to access important material properties, such as the exchange constant, at high precision.\cite{Langer2016} 

\begin{figure}[H]
\centering
\includegraphics[width=0.95\linewidth]{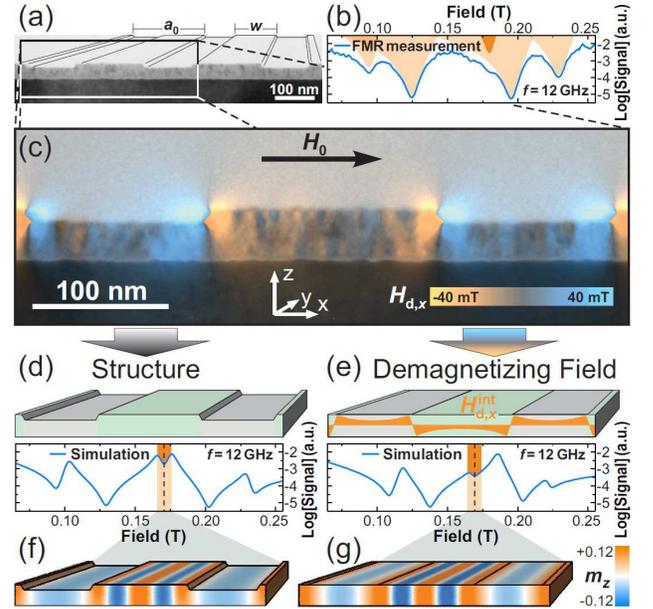}
\caption{\label{motiv} (Color online) The strategy to understand the spin-wave dynamics of a magnonic crystal. (a) Cross-sectional TEM image with (b) the measured FMR-response (Lorentzian fits in orange). (c) Magnification (black and white) superimposed with the $x$-component of the simulated demagnetizing field (colorplot). (d,e) Both properties are used as input for dynamic response simulations with (f,g) the resulting mode profiles of the highlighted resonance peaks.}
%(c) A simulation of the magnetic phase represented similarly to (b) based on the abstracted structural shape of the structure shown in (e) with constant thickness (in beam direction) as well as (d) a simulation considering a thickness distribution according to (e): The thickness variations are implemented by scaling of $M_{\mathrm{S}}$ with the same percentage as the local thickness in relation to the average thickness of $t_{\mathrm{avg}}=38.3$~nm.}
\end{figure}

In order to investigate the effect of the internal demagnetizing field $H^{\mathrm{int}}_{\mathrm{d}}$ on the spin-wave properties, the dynamics of a surface-modulated magnonic crystal (SMMC)\cite{Barsukov2011,Landeros2012,Gallardo2014} were reconstructed from `bottom-up' as sketched in Fig.~\ref{motiv}. This means, the structural shape as well as the projected demagnetizing field were mapped on the nanoscale (see Fig.~\ref{motiv}(c)) via high-resolution magnetic imaging using electron holography--a phase retrieval method in transmission electron microscopy (TEM).\cite{Lichte2008}
The results were used to reconstruct the dynamic eigenmodes of the system employing micromagnetic simulations (Fig.~\ref{motiv}(d,e)).
Comparing the results with the measurement (Fig.~\ref{motiv}(b)) yields the corresponding spin-wave states (Fig.~\ref{motiv}(f,g)) and allows to assess the role of the internal field landscape for the dynamics of the magnonic crystal.
%As the unique properties of MCs are intensively studied on the macroscale, this work intends to reconstruct the effective dynamic properties of an MC from bottom up. In that way, the effect of the internal demagnetizing field on the dynamic eigenmodes of the magnonic crystal are investigated. Therefore, the approach illustrated in Fig.~\ref{motiv} was followed. First, structural shape and the internal demagnetizing field are measured on the nanoscale (see Fig.~\ref{motiv}(a,c)) and further used in theoretical calculations as well as micromagnetic simulations (Fig.~\ref{motiv}(d-e)) to reconstruct the effective dynamic behavior of the system. Eventually, comparing the results with the measurement (Fig.~\ref{motiv}(b)) yields the corresponding spin-wave states (Fig.~\ref{motiv}(f,g)) and allows to assess the role of the internal field landscape for the dynamics of the magnonic crystal.
%
%In this context, the internal demagnetizing field $H^{\mathrm{int}}_{\mathrm{d}}$ of a magnonic crystal was mapped locally
Using ferromagnetic resonance measurements and micromagnetic simulations, the in-plane frequency-field dependence and the in-plane angular dependence of the magnonic crystal are studied together with the spin-wave mode profiles. 
%For this purpose, a 40~nm thin and \textcolor{red}{XY}~nm long lamella was prepared as a `lift-out' of a 10~nm surface-modulated MC (SMMC) of 38~nm thickness.%transmission electron microscope (TEM)
%For the TEM investigations, a 10~nm surface-modulated MC (SMMC) of 36.8~nm thickness was fabricated and from this, a 38~nm thin and 4~\textmu m long lamella was prepared.
%
%The frequency-dependent spin-wave dynamics of the MC were measured for $\textbf{k} \| \textbf{M}$ in the backward volume geometry. Furthermore, the in-plane angular dependence from backward-volume to the Damon-Eshbach geometry ($\textbf{k} \bot \textbf{M}$) was examined.

The important role of the internal demagnetizing field for the dynamics of MCs is demonstrated. Namely, it acts locally as demagnetizing \emph{and} magnetizing field. This study gains a fundamental understanding of the frequency-dependent spin-wave properties in MCs. The spin-wave behavior is examined under the rotation of the external field from the backward-volume ($\mathbf{k} \| \mathbf{M}$) to the Damon-Eshbach geometry ($\mathbf{k} \bot \mathbf{M}$) where $\mathbf{k}$ denotes the in-plane wave vector and $\mathbf{M}$ the magnetization. The angular dependence is described using the internal demagnetizing field as well as the mode localization for the estimation of an effective mode anisotropy.
%Interesting differences between SMMCs and flat (bi-component) MCs are found when following the spin-wave modes in their transition from the backward-volume ($\textbf{k} \| \textbf{M}$) to the Damon-Eshbach geometry ($\textbf{k}\bot \textbf{M}$). 
%In a broad angular range around the Damon-Eshbach geometry, spin-wave propagation was observed and is connected with a non-reciprocal dispersion.
%
\section{Theory}
\label{theo}
%
%This section introduces the fundamentals of (i) the quasi-analytical theory based on the plane wave method\cite{Sokolovskyy2011,Klos2012,Krawczyk2013} (PWM) as well as (ii) different procedures to obtain the static and dynamic properties from micromagnetic simulations.
%
The spin-wave dispersion\cite{Kalinikos1986} is the theoretical fundament for the frequency- and angle-dependent resonance equation, which is equivalently formulated in Ref.\ [\onlinecite{Gallardo2014}]:
%The theoretical fundament of the spin-wave dispersion\cite{Kalinikos1986} and angle-dependent spin-wave resonance is provided by the resonance equation as formulated in Ref.\ [\onlinecite{Kalinikos1986}]:
%\subsection{Plane Wave Method}
%
%\begin{equation}
%\left(\frac{\omega}{\gamma}\right)^{\negmedspace2}\negthickspace=\!\left( \mu_{0} H_0 + D k^2 \right) \times \left[ \mu_{0} H_0 + D k^2 + \mu_{0} M_{\mathrm{eff}} F_{pp}(kd) \right]
%\label{BVdisp}
%\end{equation}
%
%\begin{equation}
%\left(\frac{\omega}{\gamma}\right)^{\negmedspace2}=H_\mathrm{Y}(k)\cdot H_\mathrm{Z}(k)
%\label{BVdisp}
%\end{equation}
%with the stiffness fields
%\begin{align}
%\negthickspace H_\mathrm{Y}(k) \negmedspace &= \negmedspace \mu_{0} H_0 \negmedspace + \negmedspace \mu_{0} M_{\mathrm{S}} \negmedspace \left(\negthickspace %1\!-\!\frac{\left(1\!-\!\mathrm{e}^{\negmedspace-kd}\right)}{kd}\negmedspace \right)\negmedspace \sin^2\!\varphi_k \negmedspace + \negmedspace D k^2\label{Hx}\\
%\negthickspace H_\mathrm{Z}(k) \negmedspace &= \negmedspace \mu_{0} H_0 \negmedspace + \negmedspace \mu_{0} M_{\mathrm{S}}\frac{1-\mathrm{e}^{-kd}}{kd} \negmedspace + \negmedspace D k^2\label{Hy}
%\end{align}
%
\begin{equation}
\left(\frac{\omega}{\gamma}\right)^{\negmedspace2}=H_\mathrm{Y}(k)\cdot H_\mathrm{Z}(k)
\label{BVdisp}
\end{equation}
with the stiffness fields
\begin{align}
& H_\mathrm{Y}(k) = \mu_{0} H_0 + \mu_{0} M_{\mathrm{S}} \left[ 1-F(kd) \right] \sin^2\varphi_k + D k^2\label{Hx}\\
& H_\mathrm{Z}(k) = \mu_{0} H_0 + \mu_{0} M_{\mathrm{S}}F(kd) + D k^2\label{Hy}
\end{align}
Here, $f=\omega/(2\pi)$ is the spin-wave frequency, $\gamma$ the gyromagnetic ratio, $H_0$ the external magnetic field, $M_{\mathrm{S}}$ the saturation magnetization, $D=2A/M_{\mathrm{S}}$ the exchange stiffness with $A$ being the exchange constant and $\varphi_k$ is the angle between magnetization and the wave vector $\mathbf{k}$. The latter is quantized when tiny periodic thickness variations ($\Delta d << d$) are present at the surface. This circumstance leads to the occurrence of standing spin-wave modes with the wave vectors $k = 2\pi n/ a_0$ and $n = 1, 2,...$ where $a_0$ is the patterning periodicity. The term $F(kd)=[1-\mathrm{exp}(-kd)]/(kd)$ in Eqs.~(\ref{Hx}),(\ref{Hy}) is derived from the dipolar interaction with $d$ being the film thickness.
%For surface-patterned films, the wave vector of standing spin-wave modes is quantized according to $k = 2\pi n/ a_0$ with $n = 1, 2,...$ and $a_0$ the patterning periodicity. 
Note that the Eqs.~(\ref{BVdisp})--(\ref{Hy}) are only valid in the continuous thin film limit and do not account for any coupling of different spin-wave states. Furthermore, a constant magnetization profile with parallel alignment relative to the in-plane oriented external field $\mu_{0} H_0$ is assumed. However, it will be shown in Sec.~\ref{ang} that Eqs.~(\ref{BVdisp})--(\ref{Hy}) provide a reasonable estimation of the spin-wave resonance fields if the investigated spin waves branches are (i) far away from crossing points in the $f(H_0)$ dependence (such that effects of mode-coupling can be neglected) and (ii) if the film thickness $d$ and the internal demagnetizing field $H^{\mathrm{int}}_{\mathrm{d}}$ at the excited location of the structure are known. 
%Especially in the backward-volume geometry, the local demagnetizing field plays an import role. Similar to anisotropy fields, the $H_{\mathrm{d}}$ field needs to be considered for an appropriate calculation of the frequency dependence.
%
%$k = 2\pi n/ a_0$ the spin-wave wavenumber with $n = 1, 2, ...$ and $a_0$ the periodicity
\begin{figure}[t]
\includegraphics[width=0.85\linewidth]{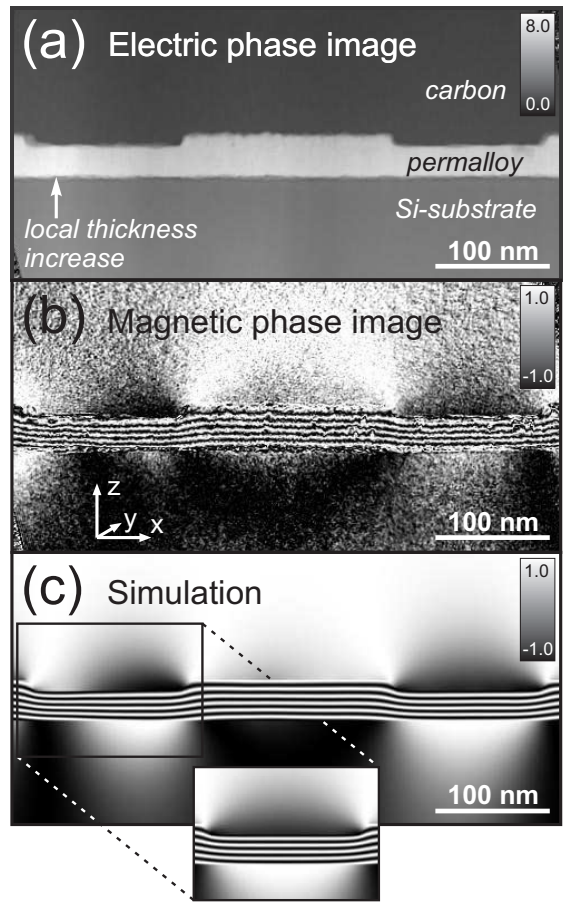}
\caption{\label{phase} (a) Electric phase image of the cross-section of a surface-modulated magnonic crystal (SMMC) with 10~nm modulation height. The contrast is proportional to the lamella thickness $t$ in $y$-direction. (b) Cosine of the $20$ times amplified magnetic phase indicating the field lines of the projected $x$,$z$ components of $\mathbf{B}$. (c) Simulation of the magnetic phase considering thickness variations (see appendices (ii,iii)). The inset depicts the simulated magnetic phase of the marked area without variations assuming a constant thickness of $t_{\mathrm{avg}}=38.3$~nm instead.}
%(c) A simulation of the magnetic phase represented similarly to (b) based on the abstracted structural shape of the structure shown in (e) with constant thickness (in beam direction) as well as (d) a simulation considering a thickness distribution according to (e): The thickness variations are implemented by scaling of $M_{\mathrm{S}}$ with the same percentage as the local thickness in relation to the average thickness of $t_{\mathrm{avg}}=38.3$~nm.}
\end{figure}

%In surface-modulated systems, internal demagnetizing fields become sufficiently strong and need to be regarded in Eq.~(\ref{BVdisp}). 
%Therefore, in a geometry around $\textbf{k} \| \textbf{M}$ for backward-volume-type spin-waves, $\mu_{0} H_0$ shall be replaced in Eqs.~(\ref{BVdisp}--\ref{Hy}) by the term $\mu_{0} H_0 + \mu_{0} H^{\mathrm{BV}}_{\mathrm{d},x} \left|\cos\varphi_k\right|$ with $H^{\mathrm{BV}}_{\mathrm{d},x}$ the maximum internal demagnetizing field in backward-volume geometry. Note, that the demagnetizing field in SMMCs has an alternating shape, and thus, changes sign between the thick and the thin part of the structure resulting in major shifts of the spin-wave resonance field depending on the mode localization.
Furthermore, this research employs semi-analytical calculations of the frequency-dependent spin-wave properties based on the plane wave method (PWM)\cite{Sokolovskyy2011,Klos2012,Krawczyk2013} with details provided in Ref.~[\onlinecite{Gallardo2016}].
%\todo{Rodolfo's playground -- `Short' version of PWM.}
%
\section{Micromagnetic Simulations}
\label{Sim}
Two kinds of micromagnetic simulations were performed as part of this research, (i) \emph{static} simulations and (ii) simulations of the \emph{dynamic} FMR response. They were calculated using the \emph{MuMax$^3$}-code\cite{Vansteenkiste2014} in two different ways, either by continuous wave\cite{Wagner2015} or by pulsed\cite{McMichael2005} excitation. 
%Generally, both approaches show very similar results, as can be seen in Sec.~\ref{res}. %However, since the time consumption is very different for both, the CW (pulsed) simulations were applied to the angle- (frequency-) dependent problems.
Further details can be found in the appendices (iv,v).
\section{Experimental Details}
\label{exp}
%
%This section is organized in the following way. At first, (i) the fabrication of the investigated MC is described, followed by (ii) the preparation of a TEM lamella and (iii) the EH analysis yielding electric and magnetic phase images and subsequently, the distribution of the $H_{\mathrm{d}}$-field. (iv) The magnetic characterization will be described in the end of the section.
%
\subsection{Sample Fabrication}
\label{Fab}
%
%The MC under investigation is a 10~nm surface-modulated magnonic crystal (SMMC) made from electron-beam-physical-vapor-deposited polycrystalline permalloy (Ni$_{80}$Fe$_{20}$). Since the fabrication method is suited to adjust and measure the modulation height, the sample has been prepared by Ar-ion milling of a $d=36.8$~nm thin film as described in ref.~[\onlinecite{Langer2016}]. To achieve an alternating film thickness, the surface of the thin film was pre-patterned by electron beam lithography (EBL). Here, ma-N 2401 negative resist was employed and structured into a wire mask with $a_0 = 300$~nm periodicity and an individual nominal stripe width of $w = 166$~nm.
Initially, a polycrystalline $d=36.8$~nm thin permalloy (Ni$_{80}$Fe$_{20}$) film was deposited on a surface-oxidized silicon substrate by electron beam physical vapor deposition. 
To achieve an alternating film thickness, the surface was pre-patterned by means of electron beam lithography. Here, ma-N~2401 negative resist was employed and structured into a stripe mask with $a_0 = 300$~nm periodicity and an individual nominal stripe width of $w = 166$~nm. Subsequently, the uncovered magnetic material was exposed to Ar-ion milling and 10~nm of the magnetic material were removed.\cite{Langer2016} In this manuscript, the resulting structure is referred to as a surface-modulated magnonic crystal.
\subsection{Electron Holography}
\label{EH}
Off-axis electron holography\cite{Lehmann2002} was employed as a unique technique to quantitatively map the projected magnetic induction\cite{Koerner2014} at a spatial resolution of about 2~nm and a magnetic phase signal resolution of about 2$\pi$/100~rad. The imaging of the 
%4.2~\textmu m $\times$ 40~nm $\times$ 1.5~\textmu m sized 
lamella was carried out in remanence using a HITACHI HF3300 (I2TEM) transmission electron microscope with a 300~kV cold field emission gun and two goniometer stages (a field-free Lorentz stage and a standard high resolution stage). 
%The lamella was saturated by the 2~T immersion field of the objective lens and the setup takes advantage of an CEOS aplanator `B-COR'\cite{Mueller}.

%
\begin{figure}[th]
\includegraphics[width=0.9\linewidth]{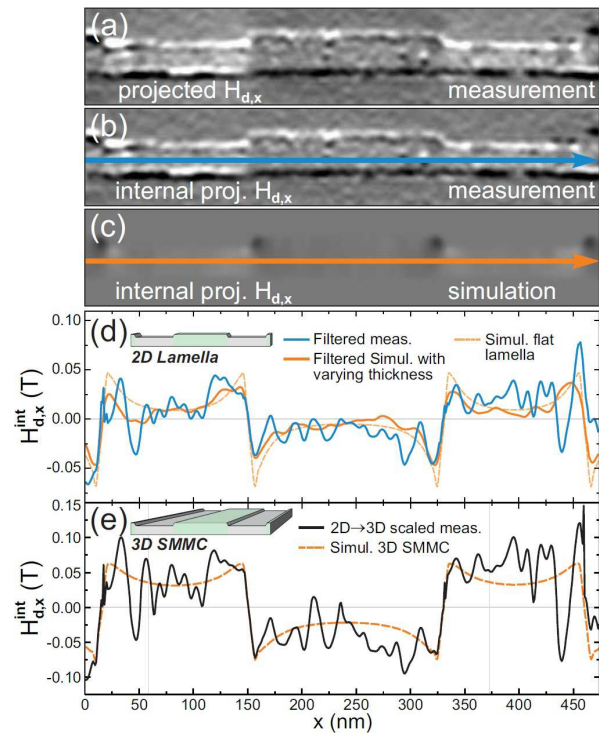}
\caption{\label{Hd} (Color online) (a) Experimentally projected $x$-component of the demagnetizing field $H_{\mathrm{d},x}$ smoothed using a Gaussian filter. (b) Same as (a) but with external $H_{\mathrm{d},x}$ field projection subtracted. (c) Simulation of the internal $H_{\mathrm{d},x}$ field. (d) The vertically averaged internal $H_{\mathrm{d},x}$ field obtained by measurement [blue arrow in (b)] and simulation [orange arrow in (c)] normalized by the thickness profile. (e) Due to the small thickness of the lamella, the internal $H_{\mathrm{d},x}$ field is scaled in order to resemble a 3D structure with continuous magnetic material along the $y$-axis (see appendix (iii) for details).}
\end{figure}
%
%Equipped with a CEOS aplanator `B-COR', the objective lens provides the field to initially saturate the lamella in its long ($x$-)direction at a 30$^{\circ}$ tilt with respect to the $y$-axis.
%The separation of the electric and magnetic phase was performed by a second holographic measurement with the sample flipped up-side down.\cite{Tonomura1986} More details regarding the measurement technique are provided in the appendix. Furthermore, information regarding the setup can be found in Refs.~[\new{add ref.}].
%
Dedicated Lorentz modes combined with the CEOS B-cor corrector allow to achieve a 0.5~nm spatial resolution in a field-free environment (less than 0.1~mT). All holograms were recorded in double-biprism configuration\cite{Harada2004} to avoid Fresnel fringes and to independently set the interface area and the fringe spacing. 
%The objective lens provides the magnetic field to initially saturate the lamella in its long ($x$-) direction at 30$^{\circ}$ tilt with respect to the beam ($y$-) direction. The separation of the electric and magnetic phase was performed by a second holographic measurement with the sample flipped up-side down.\cite{Tonomura1986} 
At a tilt of 30$^\circ$ of the lamella's long direction ($x$) with respect to the optical axis ($y$), the sample was initially saturated by means of the objective lens field. More details regarding the measurement technique are provided in the appendix (ii). Furthermore, information regarding the setup can be found in Ref.~[\onlinecite{Snoeck2014}].

The resulting electric and magnetic phase images are illustrated in Fig.~\ref{phase}(a-b). The electric phase is sensitive to different materials as well as the thickness along the beam axis. 
The magnetic phase (here: the 20 times amplified cosine) depicted in Fig.~\ref{phase}(b) appears as black and white lines reflecting the local orientation of the projected in-plane $\mathbf{B}$ field with the absolute gradient being proportional to its magnitude.

\begin{figure*}[th]
\includegraphics[width=\linewidth]{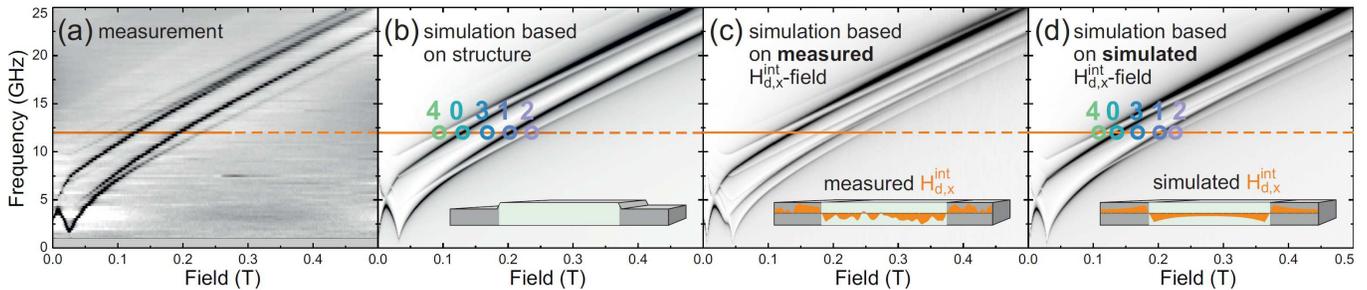}
\caption{\label{lamellafH} (Color online) $f(H_0)$ dependences of the surface-modulated magnonic crystal with the gray scale representing the dynamic response. (a) Measurement and (b-d) different simulations of the FMR response. (b) Simulation based on the structural shape of the magnonic crystal. In (c-d) the internal demagnetizing field $H^{\mathrm{int}}_{\mathrm{d},x}$ was added to a 36~nm thin film with subsequent calculation of the FMR response. (c) is based on the measured $H^{\mathrm{int}}_{\mathrm{d},x}$ field and (d) on the simulated one.}
%(c) and (d) The $f(H_0)$-dependence simulated micromagnetically by adding up (c) the measured $H_{\mathrm{d},x}$-field and (d) the simulated $H_{\mathrm{d},x}$-field to a continuous thin film.}
\end{figure*}
As explained in the appendices (ii,iii), the local thickness distribution in beam direction (plotted in Fig.~\ref{thickness}--appendix) could be obtained from the electric phase. Together with the 2D structural information according to Fig.~\ref{phase}(a), both were employed to reconstruct the remanent state of the lamella via static micromagnetic simulations.\cite{Vansteenkiste2014} The resulting simulated magnetic phase is shown in Fig.~\ref{phase}(c). Useful for comparison is an additional simulation where the thickness variations were not considered, as shown in the inset of Fig~\ref{phase}(c). 

The simulation exhibits very similar features compared to those obtained by the measurement. The phase irregularities mapped on the left side of the structure were reproduced by the simulation in Fig.~\ref{phase}(c) and can be related to a local thickness increase, since the features vanish (see inset in \ref{phase}(c)) when a flat lamella is assumed.

Next, the vertically averaged distribution of the internal demagnetizing field $H^{\mathrm{int}}_{\mathrm{d},x}(x)$ along the lamella's long axis was extracted. To achieve this, the major contribution of the magnetization was gained and subtracted from the measured projected magnetic phase by reemploying the above mentioned static simulations. Figure~\ref{Hd}(a) illustrates the resulting 2D distribution of the magnetic phase generated by the $H_{\mathrm{d},x}$ field with white (black) color representing a positive (negative) sign of $H_{\mathrm{d},x}$. This means that $H_{\mathrm{d},x}$ is acting as \emph{demagnetizing} or \emph{magnetizing} field, respectively. Figure~\ref{Hd}(b) only depicts the projected internal $H_{\mathrm{d},x}$ field isolated by subtraction of the contrast generated by the simulated stray field \emph{outside} the magnetic structure. Hence, the stray field features above the SMMC in Fig.~\ref{Hd}(a) vanish in \ref{Hd}(b). Moreover, there are parasitic contribution to the projected internal field due to the external strayfield in the front and in the back of the lamella with respect to the beam direction. Such contributions were also estimated using the static micromagnetic simulations. 
%In that way, the contributions in the front and in the back of the lamella ($y$-direction) are also considered which cannot be done by means of electron holography without additional knowledge. 

In Fig.~\ref{Hd}(d) the vertically averaged profiles of $H^{\mathrm{int}}_{\mathrm{d},x}$ are presented according to the arrows in Figs.~\ref{Hd}(b-c). The values taken from Figs.~\ref{Hd}(b-c) are normalized by the lamella thickness profile discussed in the appendix (iii). Blue color represents the measurement and orange color the simulation. The profiles demonstrate a very good agreement between both the measurement and the simulation and are corroborated with the simulated $H^{\mathrm{int}}_{\mathrm{d},x}(x)$-distribution of a flat structure (orange dashed line), where the lamella thickness was fixed to $t_{\mathrm{avg}}$.

Since further investigations focus on a 3D extended MC and not on a thin (2D) lamella structure, systematic difference between the internal fields of both systems need to be considered. As further described in the appendix (iii), this circumstance is addressed by a scaling of $H^{\mathrm{int}}_{\mathrm{d},x}(x)$ with the result shown as black solid line in Fig.~\ref{Hd}(e). Apart from the apparent oscillations arising from measurement noise, the result matches well the distribution obtained by the simulation (orange dashed line in Fig.~\ref{Hd}(e)) of an ideal 3D SMMC.
%
%The orange dashed line indicates the simulated field distribution of an ideal 3D SMMC matching well the distribution obtained by the measurement (black solid line in Fig.~\ref{Hd}(e).
%Since the small thickness of the lamella as a quasi-2D structure neglects contributions to the demagnetizing field by neighboring spins, the measured distribution requires a scaling according to the 3D structure (SMMC).
%The scaling function was obtained by a simulation of the internal $H_{\mathrm{d}}$-field distribution of a flat (2D) lamella structure (dashed line in Fig.~\ref{Hd}(c)) and an extended (3D) SMMC (dashed line in Fig.~\ref{Hd}(d)). Further details concerning this matter are given in Sec.~\ref{app}. The final result is represented by the black solid line in Fig.~\ref{Hd}(e) confirming the distribution obtained from micromagnetic simulations (orange dashed line).
%
\subsection{Magnetic Characterization}
The magnetic characterization was carried out using a broadband vector network analyzer ferromagnetic resonance (FMR) setup as described in Refs.~[\onlinecite{Langer2016},\onlinecite{Koerner2013}]. Excitation of the spin system is achieved by coupling a microwave signal via a coplanar waveguide to the surface of a `flip-chip'-mounted sample. The transmission signal $S_{21}$ is measured at several fixed excitation frequencies $f$ sweeping the external field $H_0$. The absolute value of $S_{21}$ was recorded as the FMR-response.
%
%The experiments are based on a single polycrystalline $t = 36.8$~nm thick permalloy (Ni$_{80}$Fe$_{20}$) film deposited by electron beam physical vapor deposition (EBPVD) on surface-oxidized Si(100) substrate. The surface of the film was lithographically stripe patterned using an ma-N~2001 negative resist. In order to stepwise remove the magnetic material between the resist-covered stripes [Fig.~\ref{experiment}(c-d)], sequential Ar-ion-milling was employed. The procedure is schematically depicted in Fig~\ref{experiment}(a) together with one example of a resulting SMMC. For each ion-milling step, the frequency-field-dependence $f(H_0)$ is all-electrically measured using a broadband ferromagnetic resonance (FMR) setup, as explicitly described in reference~[\onlinecite{Langer2016}].
%The etching depths of each SMMC was determined by the fitting of the $f(H_0)$ of SW modes standing vertically in the film, also referred to as perpendicular standing spin-wave (PSSW) modes. The dispersion of such modes is strongly dependent on the local film thickness and can be used to estimate the surface modulation as described in Sec.~\ref{app}.
%
\section{Results and Discussion}
\label{res}
%
%As the crucial role of the distribution of the internal demagnetizing field is already lined out in Sec.~\ref{BV}, this chapter shall focus on the relation between the internal demagnetizing field and the dispersion and mode profiles of hybrid SWs in SMMCs. Therefore, a transmission electron microscopy (TEM)-holography analysis of an SMMC with 10~nm surface modulation was conducted employing the same permalloy film as used in Sec.~\ref{BV}. The method allows for a screening of the local external and internal strayfield of the structure and yields at the same time high-quality images of the structural properties. Employing numerical simulations, both approaches can then be followed to calculate the FMR response: (i) Using the structural properties on the one hand and (ii) adding the measured internal $H_{\mathrm{d}}$-field distribution to a regular permalloy thin film, similar to the approach of the analytical theory, on the other hand.
%
In this section, the results from two independent approaches to reconstruct the effective spin-dynamics in a magnonic crystal are discussed. With both the knowledge of (i) the structural shape and (ii) the internal $H_{\mathrm{d},x}$ field, dynamic simulations were performed. 
%The former (i) is used to remodel structurally the SMMC and simulate the FMR response whereas the latter (ii) is virtually added as an internal field to a continuous film with subsequent dynamic response simulations.
%
\subsection{Frequency dependence in backward-volume geometry}
\label{frequ}
Figure.~\ref{lamellafH} illustrates several $f(H_0)$ dependences obtained from measurement and simulations. In \ref{lamellafH}(a) the measured $f(H_0)$ is shown whereas \ref{lamellafH}(b) was obtained from the remodeling of the sample structure and subsequent FMR simulations. The evident similarity between both indicates a reliable representation of the sample structure by the micromagnetic model. In contrast, Fig.~\ref{lamellafH}(c) and (d) are obtained by simulating a 36~nm thin permalloy film with a virtually added periodic distribution of $H^{\mathrm{int}}_{\mathrm{d},x}$. In \ref{lamellafH}(c) the \emph{measured} field distribution was employed and in \ref{lamellafH}(d) the \emph{simulated} one was taken corresponding both to the two plots in Fig.~\ref{Hd}(e). A convincing qualitative agreement of all shown $f(H_0)$ dependences with the measurement is obtained.

However, at second glance, a higher number of modes can be found in Fig.~\ref{lamellafH}(c), which is due to the measurement noise in $H^{\mathrm{int}}_{\mathrm{d},x}$ violating the mirror symmetry of the internal field landscape. Especially at the edges of the thick part, the different local demagnetizing fields lead to the occurrence of two separate \emph{non-symmetric} edge modes with different energies.
However, for the symmetric $H^{\mathrm{int}}_{\mathrm{d}}$ field in Fig.~\ref{Hd}(d), the $f(H_0)$ matches well the one obtained in Fig.~\ref{Hd}(b) with similar mode characteristics.
\subsection{Mode profiles}
\label{prof}
Another way to test the level of similarity between the different simulations presented above is to analyze the mode profiles. In Fig.~\ref{profiles12GHz} the profiles of the resonant spin waves at $f=12$~GHz are plotted and labeled with the respective mode number $n$. The plots indicate a convincing agreement between both simulations such that the individual character of the plotted mode profiles reflects similar physics. Consequently, the dynamics of the SMMC is very similar to flat MC with a pronounced internal field structure, such as bi-component MCs. The dynamics of such systems can be well described by the plane wave method,\cite{Sokolovskyy2011,Klos2012,Krawczyk2013,Gallardo2016} which was used in addition for the calculation of the mode profiles in Fig.~\ref{profiles12GHz} confirming the results from the simulations. Note that the frequency of 12~GHz was selected such that effects from mode coupling are small and, thus, can be neglected in the following discussion.
%Similar to bi-component systems, the mode profiles of spin-waves in an SMMC can be described semi-analytically using the plane wave method\cite{Sokolovskyy2011,Klos2012,Krawczyk2013}. The resulting mode profiles are additionally plotted in Fig.~\ref{profiles12GHz} with similar results. Thus, a governing role of the dynamics of the MC by the internal demagnetizing field is concluded.

\begin{figure}[t]
\includegraphics[width=1\linewidth]{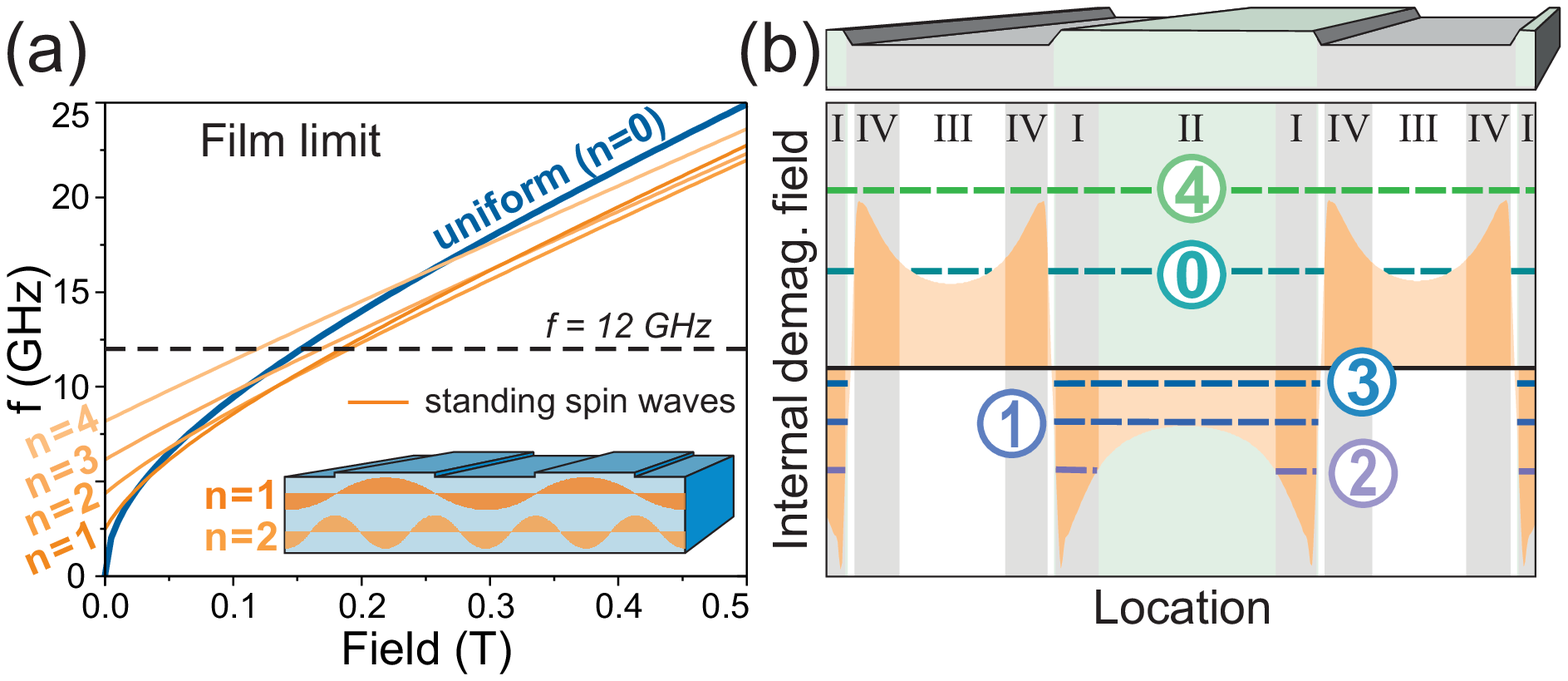}
\caption{\label{SSW} (Color online) (a) $f(H_0)$ in the film limit with tiny modulation. Standing spin-waves modes appear quantized due to $k = 2\pi n/ a_0$ and are sketched in the inset.
%At $f=12$~GHz (dashed line), spin waves with $n=1,2$ and 3 are energetically below the uniform one. 
(b) Mode localization and internal field landscape of an SMMC. The $H^{\mathrm{int}}_{\mathrm{d}}$ field is negative in region I and II and positive in region III and IV. Thus, modes with energy below the uniform mode (flat black line) can only be excited in the regions where the internal field is reduced (I,II).}
%The local energy is defined by the $H^{\mathrm{int}}_{\mathrm{d},x}$ field, i.e.\ lower modes tend to localize where $H^{\mathrm{int}}_{\mathrm{d},x}$ minimizes. Thus, the $n=1,2,3$ modes are localized in the thick part (I,II) whereas the others extend into the thin part (III,IV).}
\end{figure}

The reason for the multitude of measurable spin-wave modes in SMMCs is explained by Fig.~\ref{SSW}(a). Here, the $f(H_0)$ dependence of the modes in the limit of a thin film with tiny modulation $\Delta d\rightarrow 0$ is plotted. Apart from the uniform mode, standing spin-wave modes are present, with a defined number of nodes ($2n$) fitting in one period $a_0$ as sketched in the inset of Fig.~\ref{SSW}(a). This circumstance results in a quantization of the wave vector with $k = 2\pi n/ a_0$. By applying Eqs.~(\ref{BVdisp})--(\ref{Hy}), the corresponding frequency dependence can be obtained (orange lines). The standing spin-wave modes can couple to the uniform mode and together form the full spectrum of possible states accessible in such structures.\cite{Landeros2012,Gallardo2014,Langer2016} In Fig.~\ref{SSW}(a), at the marked frequency of 12~GHz, three states with lower energy than the uniform one with $n=1,2,3$ are found and with the $n=2$ state being lowest. Note that for a given frequency, the mode energy is reflected by the resonance field such that for low (high) energy modes a high (low) external field must be supplied to resonate at the same frequency. Thus, at $f=\mathrm{const.}$, high resonance fields represent low mode energy and reverse.

For an SMMC with a pronounced modulation, these states are present as well, but are shaped differently by the internal field landscape. In Fig.~\ref{profiles12GHz}, all modes can still be identified according to their total number of nodes ($2n$) inside a period $a_0$. However, due to the presence of the internal field landscape, the modes can no longer be described assuming a constant wave vector due to $k=2\pi n/a_0$ and an extension over the full MC. Instead, the modes 0--3 reveal a clear localization in either the thick or the thin part and all modes show major deviations from the harmonic character sketched in the inset of Fig.~\ref{SSW}(a), which can only be explained with the help of the internal field landscape shown in Fig.~\ref{SSW}(b). The field distribution (orange) is translated into a region map (roman numbers) of negative (I,II) and positive (III,IV) internal fields. The dashed lines represent the part of the field landscape where the respective mode energy is sufficient for a spin-wave excitation.

In order to understand the characteristic mode profiles in Fig.~\ref{profiles12GHz}, it is useful to know the dependence of the wave vector $k$ on the effective field $H^n_{\mathrm{eff}}=H^n_0+H^{\mathrm{int}}_{\mathrm{d}}$. At this point, the knowledge of the internal demagnetizing field $H^{\mathrm{int}}_{\mathrm{d}}$ becomes relevant again. As the $H^{\mathrm{int}}_{\mathrm{d}}$ field itself depends on the location along the $x$-axis, the distribution $H^{\mathrm{int}}_{\mathrm{d}}(x)$ can be used to assign a specific $k$-value with a location inside the MC. Moreover, this relation can be used to identify regions where no $k$-value can be attributed to the effective field which is important for understanding the individual mode localization.
For this purpose, the spin-wave dispersion expressed by Eqs.~(\ref{BVdisp})--(\ref{Hy}) is employed with $H_0$ being replaced by the effective field $H^n_{\mathrm{eff}}=H^n_0+H^{\mathrm{int}}_{\mathrm{d}}$ to consider both, the external field of the $n$th spin wave in resonance $H^n_0$ as well as the internal demagnetizing field $H^{\mathrm{int}}_{\mathrm{d}}$. Accordingly, the dependence of the effective field $H^n_{\mathrm{eff}}$ on the wave vector $k$ reads (for $\varphi_k=0^\circ$):
\begin{figure}[t]
\includegraphics[width=0.9\linewidth]{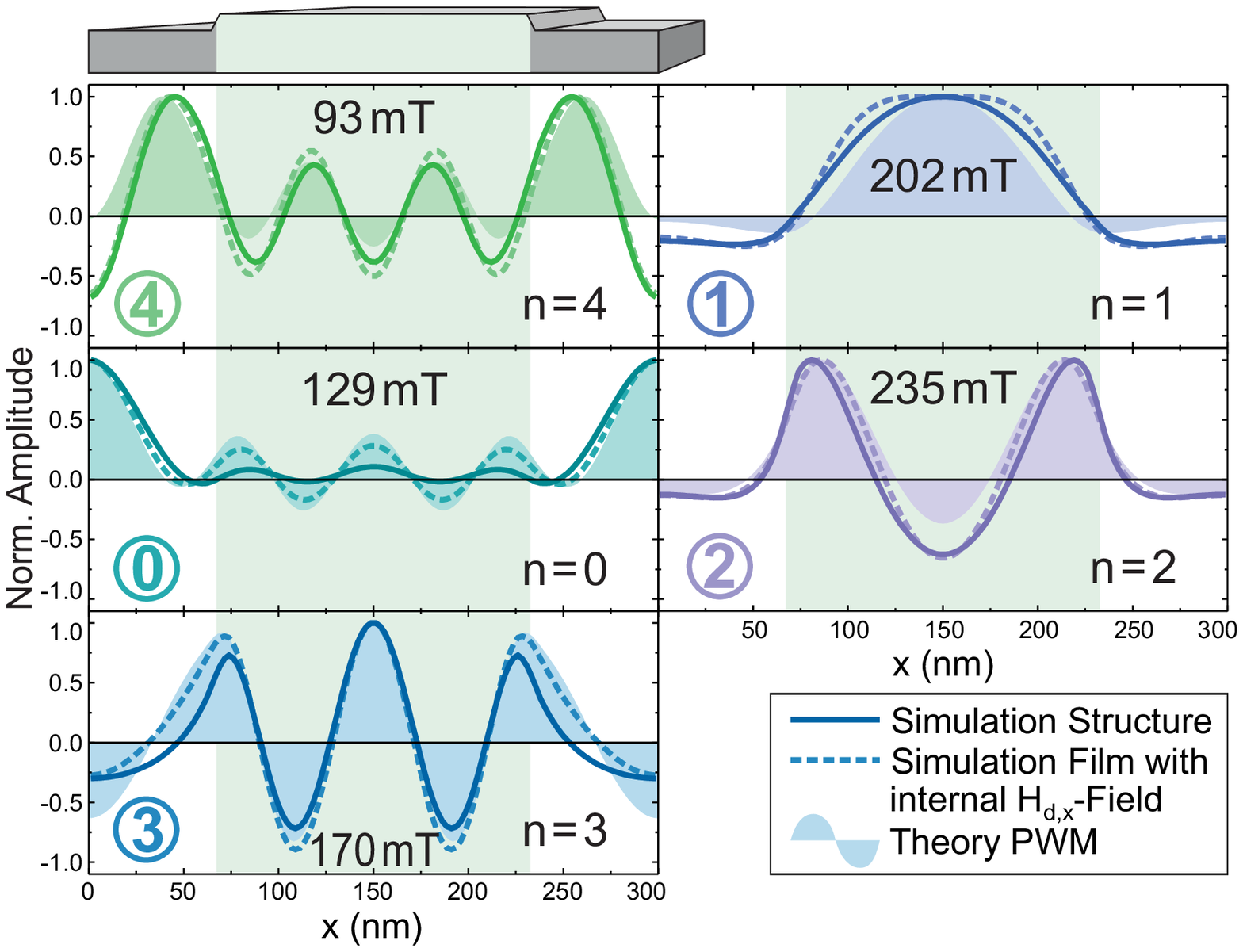}
\caption{\label{profiles12GHz} (Color online) Mode profiles of spin-waves (with corresponding mode number $n$) at different resonance positions as marked in Figs.~\ref{lamellafH} and \ref{azimuth}. They are derived by simulations and PWM theory.}
\end{figure} 
\begin{equation}
\mu_{0}H^n_{\mathrm{eff}}\!=\!-\tfrac{1}{2}\mu_{0}M_{\mathrm{S}}F\!+\negthickspace\bigg[\!\tfrac{1}{4}\!\left(\mu_{0}M_{\mathrm{S}}F\right)^{\!2}\negthickspace+\!\left(\frac{\omega}{\gamma}\right)^{\negthickspace 2} \bigg]^{\!\tfrac{1}{2}}
\label{efffield}
\end{equation}
\noindent Eq.~(\ref{efffield}) can now be used, to correlate the wave vector with the effective field at $f=12$~GHz, which is illustrated in Fig.~\ref{k}(a) for both the thick and the thin part of the MC. With the given resonance fields in Fig.~\ref{profiles12GHz} and the knowledge of the internal demagnetizing field $H^{\mathrm{int}}_{\mathrm{d}}(x)$, the effective fields can be calculated for all different locations in the SMMC and for each spin-wave mode. The colored lines in Fig.~\ref{k}(a) correspond to the range of $k$-values associated with the internal field landscape for each mode. Bright colors represent the edge regions (I,IV) and dark colors represent the center regions (II,III). In Fig.~\ref{k}(b), the $H^{\mathrm{int}}_{\mathrm{d}}(x)$ distribution (orange dashed line in Fig.~\ref{Hd}(e)) is used, to calculate the wave vector dependent on the location along the $x$-axis.

With Fig.~\ref{k}(a) and (b), the reason for the mode localization can be explained. For modes 1--3, the effective field in the thin part (III,IV) exceeds 176~mT, which is maximum value (vertex of the gray parabola in \ref{k}(a)) for a defined spin-wave excitation in this region. Thus, all three modes localize in the thick part (I,II) and avoid the regions III and IV. Moreover, the calculations reveal that mode 2 is only excited at the edges of the thick part (I). 
%Supported by the mode profile in Fig.~\ref{profiles12GHz}, this mode exhibits a tendency towards an edge mode, which might become even more prominent for a stronger surface modulation.

It is important to note that in SMMCs with a pronounced modulation, a classical uniform mode cannot exist due to the variance of the internal fields. Instead, mode 0 behaves as a quasi-uniform excitation of the center of the thin part (III,IV) of the MC, which is supported by the mode profile in Fig.~\ref{profiles12GHz} and by the range of $k$-values in Fig.~\ref{k}(b) reaching almost perfectly $k=0$ in the center of part III. Unlike the higher modes 2--4, the wave vector of mode 0 and mode 1 is not only delimited by the vertices of the parabolae in Fig.~\ref{k}(a) where the energy becomes too small for a spin-wave excitation. It is also delimited by the uniform state ($k=0$) at $\mu_0 H_{\mathrm{eff}}=154$~mT such that regions of lower internal fields cannot be excited anymore. Due to that reason, mode 0 avoids the thick part (I,II) as much as mode 1 avoids region I as shown in Fig.~\ref{k}(b) and confirmed by Fig.~\ref{profiles12GHz}.
\begin{figure}[t]
\includegraphics[width=1\linewidth]{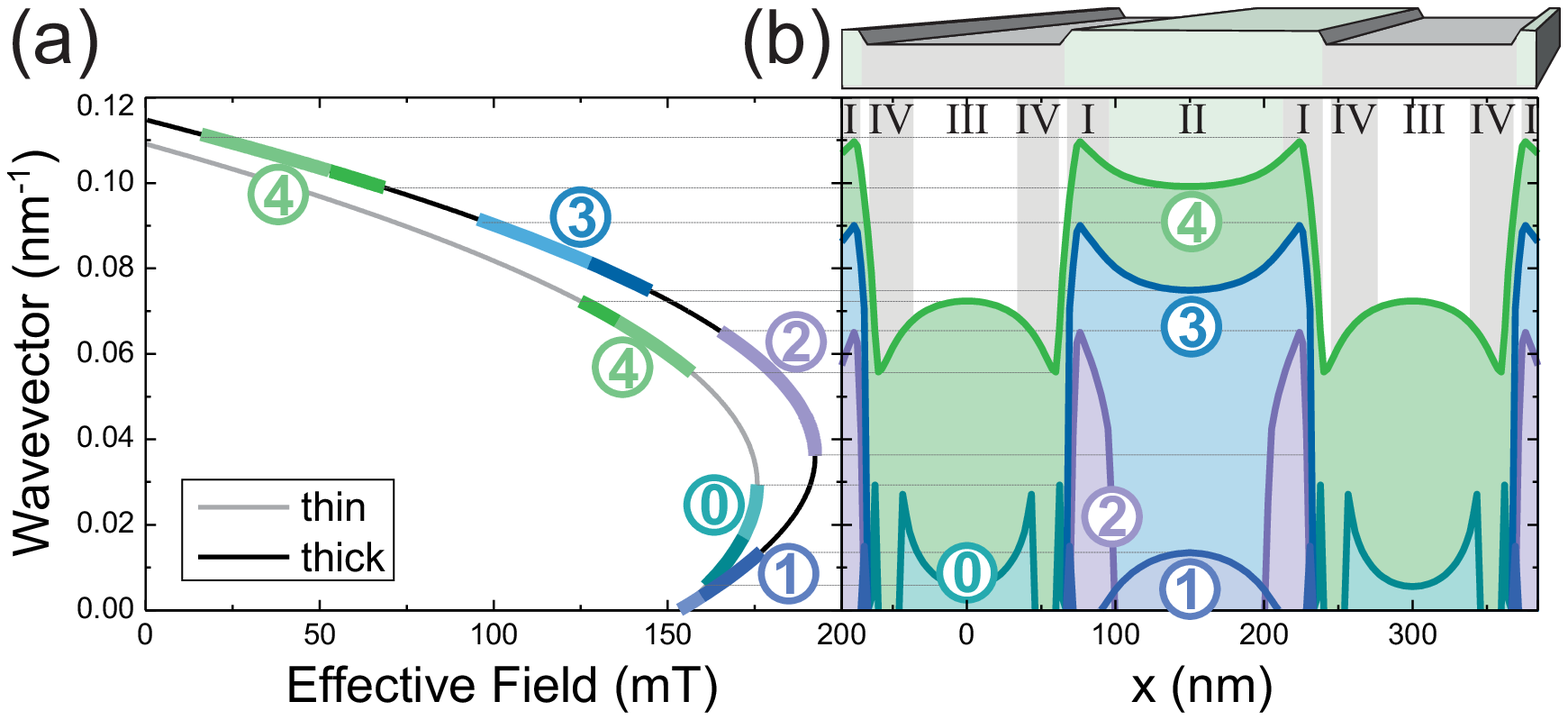}
\caption{\label{k} (Color online) wave vector calculation for the resonances marked in Fig.~\ref{profiles12GHz} dependent on (a) the effective field and (b) the location along the $x$-axis. For modes 1--3, the internal fields in the thin part are so high that the effective field exceeds the vertex of the parabola in (a) and thus, this region is avoided. In (a) bright lines correspond to the edge (I,IV) and dark lines correspond to the center regions (II,III).}
\end{figure}

The only mode with sufficient energy to spread over the full MC is mode 4. In Fig.~\ref{k}(a) and (b) the distribution of the modes' wave vector is plotted according to Eq.~(\ref{efffield}). Expressed vividly, the mode can rearrange its 8 nodes in a way that the energy of the mode is distributed equally over the full structure. The number of nodes in the thick part $m$ and in the thin part $l$ can be estimated by solving $\mu_{0}H^{\mathrm{thick}}_0=\mu_{0}H^{\mathrm{thin}}_0$, i.e.\ $\mu_{0}H^{m}_0\left( d=36\,\mathrm{nm},\mu_{0}H^{\mathrm{thick}}_{\mathrm{d}}=-39\,\mathrm{mT},k=\frac{m\pi}{w}\right)=\mu_{0}H^{l}_0\left( d=26\,\mathrm{nm},\mu_{0}H^{\mathrm{thin}}_{\mathrm{d}}=41\,\mathrm{mT},k=\frac{l\pi}{a_0-w}\right)$

\noindent with $n=m+l$ and with $H^{\mathrm{thick}}_{\mathrm{d}}$ and $H^{\mathrm{thin}}_{\mathrm{d}}$ being the average internal demagnetizing fields of the thick and the thin part of the MC. Applying Eq.~(\ref{efffield}) yields a resonance field of $\mu_{0}H^{m}_0=\mu_{0}H^{l}_0=103$~mT and the node numbers $m=5.29$ and $l=2.71$, which is coherent with the node distribution in Fig.~\ref{profiles12GHz}.

In short, it is observed, that three kinds of modes are distinguished in the SMMC. (i) A quasi-uniform central excitation of the thin part of the SMMC, which corresponds to mode 0. (ii) $k\neq0$ modes with sufficient energy to extend over the full MC (e.g.\ mode 4) and (iii) $k\neq0$ modes with insufficient energy enforcing a localization in the thick part (I,II) of the MC, such as mode 1--3.

Modes of category (ii) adapt their wave vector such that the mode energy is equally distributed over the full structure while the total number of nodes ($2n=m+l$) is conserved. For these modes, the wave vector must be calculated separately for both the thick and the thin part as explained above. This is different for the category (iii) of localized modes. These modes exhibit a `damped' trough in the thin part where the local fields are too high for a spin-wave excitation. The residual $2n-1$ nodes of the modes are condensed in the thick part, where the internal field is reduced. Accordingly, the wave vector of these modes is shifted to $k=(2n-1)\pi/w$ instead of $2\pi n/ a_0$ in the thin film limit.
\subsection{Angular Dependence}
\label{ang}
%
%Examining the angular dependence presented in Fig.~\ref{azimuth} allows to follow the spin-waves in their transition from a highly-pinned character with strong internal fields in backward-volume geometry ($\varphi=0^\circ$, 180$^\circ$) to the Damon-Eshbach geometry (90$^\circ$, 270$^\circ$) with almost negligible internal fields.

\begin{figure}[t]
\includegraphics[width=\linewidth]{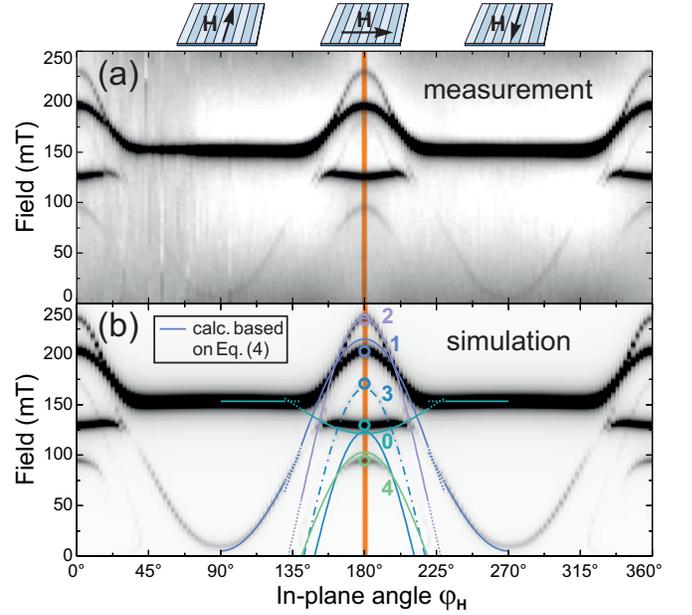}
\caption{\label{azimuth} (Color online) In-plane angular dependence of the resonance fields of an SMMC with 10~nm modulation height, where (a) is the measurement and (b) the corresponding simulation. The numbered resonances correspond to the mode profiles shown in Figs.~\ref{profiles12GHz}. The solid lines are calculations based on Eq.~(\ref{azi}) with parameters accounting for $\varphi_H=135^\circ$--225$^\circ$ provided in table~\ref{tabsw}. For $\varphi_H=225^\circ$--270$^\circ$, mode 1 was described without consideration of an internal demagnetizing field assuming a film thickness of $d=26$~nm.}
%The blue lines are analytical calculations of the $n=1$ spin-wave mode according to Eq.~(\ref{BVdisp}). Note that in backward volume geometry ($\varphi_H=0^\circ, 180^\circ$) the mode behaves similarly to a 36~nm film mode with an additional anisotropy field of -22~mT. In contrast, around the Damon-Eshbach geometry ($\varphi_H=90^\circ, 270^\circ$), the mode follows the behavior of a 26~nm film mode.}
\end{figure}

Figure~\ref{azimuth}(a) shows the measurement and \ref{azimuth}(b) the simulation of the angular dependence at $f=12$~GHz. The backward-volume direction ($\varphi_H=0^\circ$, 180$^\circ$) is marked by the orange line with the labeled resonances being the same as in Fig.~\ref{lamellafH}(b). $\varphi_H=90^\circ$ and 270$^\circ$ both correspond to the Damon-Eshbach geometry. Again, a satisfactory reconstruction of the measurement by the simulation based on the sample structure is obtained. 

The most prominent resonance branch is the flat one between 45$^\circ$--135$^\circ$ and 225$^\circ$--315$^\circ$. This mode corresponds to the uniform mode around the Damon-Eshbach geometry with negligible internal demagnetizing fields. In the same angular range, there is a second less noticeable resonance branch observed at lower external fields corresponding to the $n=1$ Damon-Eshbach mode. Knowing that the $n=1$ mode is identified at $\mu_{0}H_0=202$~mT in the backward-volume direction, mode 1 can be followed through a full 360$^\circ$ rotation of the external field. 
%For an intuitive access to Fig.~\ref{azimuth}, one needs to keep in mind, that \emph{high} resonance fields correspond to \emph{low} mode energies.

In order to analytically express the angular dependence of a mode, Eqs.~(\ref{BVdisp})--(\ref{Hy}) can again be employed together with the identity $\varphi_k=\varphi_H$.
%The approach can be applied to the above mentioned angular ranges around the Damon-Eshbach geometry, where internal fields play a minor part. Interestingly, a reliable reproduction of the behavior of mode 1 (dark blue line in Fig.~\ref{azimuth}(b)) can only be obtained if a dynamically active film thickness of only $d=26$~nm (corresponding to the thin part) is assumed. Such approach fails if the same concept is applied to the backward-volume direction.
Around the backward-volume direction, the high internal demagnetizing fields must also be taken into account with regard to the individual mode localization. In order to include the demagnetizing field into the angle-dependent spin-wave dispersion, $\mu_{0}H_0$ was replaced by $\mu_{0}H_0+\mu_{0}H_{\mathrm{d}}\cdot\cos{(2\varphi_H)}$ in Eq.~(\ref{Hx}) and by  $\mu_{0}H_0+\mu_{0}H_{\mathrm{d}}\cdot\cos^2\negmedspace{\varphi_H}$ in Eq.~(\ref{Hy}) analogous to the description of a uniaxial anisotropy field.\cite{Lenz2005,Koerner2013,Lindner2013} From Eqs.~(\ref{BVdisp})--(\ref{Hy}), a modified angular dependence is obtained
\begin{equation}
\begin{split}
&\mu_{0}H^n_0(\!\varphi_{\!H}\!)=-\tfrac{1}{2}\mu_{0}H_{\mathrm{d}}\!\left(\cos(2\varphi_{\!H}\!)\!+\!\cos^2\negthickspace{\varphi_{\!H}}\right)\!-\!Dk^2\\
&-\!\tfrac{1}{2}\mu_{0}M_{\mathrm{S}}F\!-\!\tfrac{1}{2}\mu_{0}M_{\mathrm{S}}(\!1\negmedspace -\negmedspace F)\sin^2\negthickspace{\varphi_{\!H}}\negmedspace+\negthickspace\bigg[\tfrac{1}{4}\mu^2_{0}H^2_{\mathrm{d}}\sin^4\negthickspace\varphi_{\!H}\\
&+\!\tfrac{1}{2}\mu^2_{0}H_{\mathrm{d}}M_{\mathrm{S}}F\sin^2\negthickspace{\varphi_{\!H}}\!-\!\tfrac{1}{2}\mu^2_{0}H_{\mathrm{d}}M_{\mathrm{S}}(\!1\negmedspace -\negmedspace F)\sin^4\negthickspace{\varphi_{\!H}}\\
&+\!\tfrac{1}{4}\!\left(\mu_{0}M_{\mathrm{S}}F\!-\!\mu_{0}M_{\mathrm{S}}(\!1\negmedspace -\negmedspace F)\sin^2\negthickspace{\varphi_{\!H}}\right)^2\!+\!\left(\frac{\omega}{\gamma}\right)^{\negthickspace 2} \bigg]^{\tfrac{1}{2}}
\end{split}
\label{azi}
\end{equation}
\noindent with $\mu_{0}H^n_0$ the resonance field of the $n$th mode.
% and again with the $k$-vector being quantized after $k_{\|} = 2\pi n/ a_0$. 
The angle-dependent resonance fields are calculated using Eq.~(\ref{azi}) employing simplified assumptions: (i) The wave vector of localized $k\neq0$ modes is defined by $k=(2n-1)\pi/w$ and (ii) for the effective demagnetizing field $H^{\mathrm{eff}}_{\mathrm{d}}$ the average value of the regions in which the modes localizes is taken. 
%
%the wavevector is quantized due to $k = 2\pi n/ a_0$ and (ii) the thickness $d$ is selected according to the part where the mode is localized and (iii) the spin-wave modes are exposed to an effective demagnetizing field $H^{\mathrm{eff}}_{\mathrm{d}}$, which is estimated as follows:
%
%\begin{equation}
%\mu_0 H^{\mathrm{eff}}_{\mathrm{d}}=\int\limits^{a_0}_{0} \mu_0 H^{\mathrm{int}}_{\mathrm{d}}(x)\cdot\left|\psi^n(x)\right|^2\mathrm{d}x
%\label{SWaniso}
%\end{equation}
%
%\noindent with $\psi^n(x)$ being the mode profile, which is normalized according to $\int\left|\psi^n(x)\right|^2\mathrm{d}x=1$. The resulting values of $\mu_0 H^{\mathrm{eff}}_{\mathrm{d}}$ are presented in table~\ref{tabsw}. 
%
(iii) As explained in Sec.~\ref{prof}, for modes localized in the thick as well as the thin part of the MC (e.g.\ mode 4), the node number and the effective field $H^{\mathrm{eff}}_{\mathrm{d}}$ are calculated separately for both parts.

The calculated angle dependences according to Eq.~(\ref{azi}) are depicted in Fig.~\ref{azimuth} as solid lines revealing a firm overall agreement to the measurement and the simulation. The parameters used for the calculations according to the above assumptions are provided in table~\ref{tabsw}. Mode 3 is the only one with major deviations from the resonance positions in the colorplot. The discrepancy is likely due a different pinning condition at the edge of the thick part resulting in an overestimation of the wave vector by $k=(2n-1)\pi/w$. This is supported by the mode profile in Fig.~\ref{profiles12GHz} revealing a reduced wave vector between the film limit $2\pi n/ a_0$ and $(2n-1)\pi/w$. A fitting angle-dependent resonance position can be obtained for $k=4.3\pi/w$ (blue dot-dashed line in Fig.~\ref{azimuth}(b)), which is coherent with the number of nodes in Fig.~\ref{profiles12GHz}.

For the calculations in and around the Damon-Eshbach geometry, the internal demagnetizing fields were neglected, i.e.\ , $\mu_0 H^{\mathrm{eff}}_{\mathrm{d}}=0$. Interestingly, a reliable reproduction of the behavior of mode 1 (dark blue line in Fig.~\ref{azimuth}(b)) can only be obtained if a dynamically active film thickness of only $d=26$~nm (corresponding to the thin part) is assumed.
%
%is given by the center of the corresponding part in which the mode is excited: $\mu_{0}H_{\mathrm{d}}=-22$~mT for modes ($n=1,2,3$) excited in the thick part and $\mu_{0}H_{\mathrm{d}}=+32$~mT for the modes ($n=0,4$) excited in the thin part and (iii) the thickness $d$ being defined by the excitation region. For clarification, all used parameters are provided in table~\ref{tabsw}. The angular dependence fits quite well for the modes 0, 1 and 3. In contrast, mode 2 shows systematic deviations from the measurement and the simulation.
%
%\begin{table}[b]
%\begin{center}
%\caption{Parameters used for a simple description of the spin waves' angular dependence between $\varphi_H=135^\circ$--225$^\circ$. The effective demagnetizing field $\mu_{0}H^{\mathrm{eff}}_{\mathrm{d}}$ is obtained by Eq.~(\ref{SWaniso}).}
%\label{tabsw}
%\begin{tabular}{lccccccccccc}
%\hline\hline
%mode no.\ $n$ & ~ & 0 & ~ & 1 & ~ & 2 & ~ & 3 & ~ & 4 ($n'=2.71$)\\
%\hline
%exc.\ part & ~ & thin & ~ & thick & ~ & thick & ~ & thick & ~ & thin/thick \\
%$d$~(nm) & ~ & 26 & ~ & 36 & ~ & 36 & ~ & 36 & ~ & 26 \\
%$\mu_{0}H^{\mathrm{eff}}_{\mathrm{d}}$~(mT) & ~ & 32.7 & ~ & -24.6 & ~ & -44.0 & ~ & -27.6 & ~ & 46.8 \\
%\hline\hline
%\end{tabular}
%\end{center}
%\end{table}
%
\begin{table}[b]
\begin{center}
\caption{Parameters used for the calculation angular dependence of the spin waves between $\varphi_H=135^\circ$--225$^\circ$.}
\label{tabsw}
\renewcommand{\arraystretch}{1.25}
\begin{tabular}{ccccccccc}
\hline\hline
mode no.\ $n$ & ~ & \footnotesize{localisation} & ~ & $d$~(nm) & ~ & $k_{\mathrm{eff}}$ & ~ & $\mu_{0}H^{\mathrm{eff}}_{\mathrm{d}}$~(mT) \\
\hline
0 & ~ & III & ~ & 26 & ~ & 0 & ~ & 31.9 \\
1 & ~ & II & ~ & 36 & ~ & $\pi/w$ & ~ & -31.8 \\
2 & ~ & I & ~ & 36 & ~ & $3\pi/w$ & ~ & -57.0 \\
3 & ~ & I,II & ~ & 36 & ~ & $5\pi/w$ & ~ & -38.7 \\
\multirow{2}{*}{4} & \multirow{2}{*}{~} & \multirow{2}{*}{I--IV} & ~ & 26 & ~ & $\frac{2.71\pi}{a_0-w}$ & ~ & -38.7 \\
& ~ &  & ~ & 36 & ~ & $5.29\pi/w$ & ~ & 40.9 \\
%4 & ~ & I--IV & ~ & 26 (36) & ~ & $\tfrac{2.71\pi}{w}$ ($\tfrac{2.71\pi}{w}$) & ~ & -38.7 (40.9) \\
\hline\hline
\end{tabular}
\end{center}
\end{table}
\section{Conclusion}
\label{Con}
Electron holography measurements were employed to map the internal magnetic field landscape of a surface modulated magnonic crystal on the nanoscale. The measurements confirmed the alternating character of the demagnetizing field acting locally as demagnetizing- and magnetizing field. Micromagnetic reconstructions of its dynamic behavior revealed the dominating role of the magnonic crystals' internal demagnetizing field. The significant impact of the internal field landscape on the mode profiles and the modes' angular dependences were discussed. 
\section{Acknowledgment}
We thank B.\ Scheumann for the film deposition, A.\ Kunz for the FIB lamella preparation and Y.\ Yuan and S.\ Zhou for the VSM characterization as well as H.\ Lichte for fruitful discussions. Support by the Nanofabrication Facilities Rossendorf at IBC as well as the infrastructure provided by the HZDR Department of Information Services and Computing are gratefully acknowledged. Our research has received funding from the Graduate Academy of the TU Dresden, from the European Union Seventh Framework Program under grant no.\ 312483-ESTEEM2 (Integrated Infrastructure Initiative-I3), the Centers of Excellence with Basal/CONICYT financing (grant no.\ FB0807), CONICYT PAI/ACADEMIA 79140033, FONDECYT 1161403, CONICYT PCCI (grant no.\ 140051) and DAAD PPP ALECHILE (grant no.\ 57136331) and from the Deutsche Forschungsgemeinschaft (grant no.\ {LE2443/5-1}).
\section{Appendix}
\label{app}
This section contains details regarding (i) the fabrication of the TEM lamella, (ii) the electron holography technique, (iii) the extraction of the internal demagnetizing field and (iv) the static and (v) the dynamic simulations carried out in this work.

\vspace{3mm}
\textbf{(i) Lamella Fabrication}\quad The cross-sectional TEM lamella of the magnonic crystal was prepared by in-situ lift-out using a Zeiss Crossbeam NVision 40 system. In order to protect the structure surface, a carbon cap layer was deposited by electron beam assisted precursor decomposition and subsequent Ga focused ion beam (FIB) assisted precursor decomposition. Subsequently, the TEM lamella was prepared using a 30~keV Ga FIB with adapted currents. Its transfer to a 3-post copper lift-out grid (Omniprobe) was done with a Kleindiek micromanipulator. To minimize sidewall damage, Ga ions with 5~keV energy were used for final thinning of the TEM lamella until electron transparency was achieved.
\begin{figure}[t]
\includegraphics[width=0.9\linewidth]{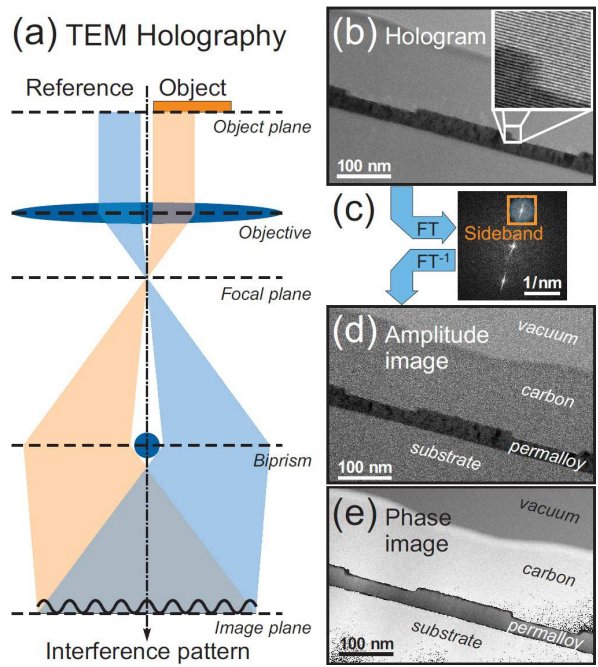}
\caption{\label{TEM} (Color online) Acquisition and reconstruction scheme of an electron hologram. (a) Setup of electron holography in TEM. (b) Hologram of a permalloy (Ni$_{80}$Fe$_{20}$) thin film with $\Delta d=10$~nm surface modulation. (c) Fourier spectrum of the hologram showing two sidebands and one center band. The Fourier transform of the upper sideband, low-pass filtered by a numerical aperture, yields the image wave represented by (d) the (wrapped) amplitude and (e) the phase image.}
\end{figure}

\vspace{3mm}
\textbf{(ii) Off-Axis Electron Holography}\quad Figure~\ref{TEM}(a) illustrates the working principle of an off-axis electron holography setup. Employing a M\"ollenstedt biprism, the object- and the reference beam is precisely superimposed at the image plane. The recorded interference fringe pattern is shown in Fig.~\ref{TEM}(b) with tiny contrast variations and fringe bending (see inset in Fig.~\ref{TEM}(b)). The hologram is reconstructed by employing the upper sideband of the hologram's Fourier spectrum (see Fig.~\ref{TEM}(c)). By inverse Fourier transformation the amplitude and phase information depicted in Fig.~\ref{TEM}(d) and Fig.~\ref{TEM}(e) are obtained. The phase unwrapping is carried out using the Goldstein algorithm.\cite{Perkes2002} The hologram series acquisition (40 holograms for each orientation) and the wave averaging were employed to reduce the phase noise.\cite{Roeder2014} Note that displacement removal and first-order aberration corrections were required to match the mean phase.

The electron phase is sensitively altered by electric and magnetic properties of the sample and is, thus, key quantity for the field mapping on the nanoscale\cite{Lichte2008} given by
\begin{equation}
\varphi(x,z) = C_E \int\limits_{t_l(x,z)}^{t_u(x,z)} V(x,y,z)\mathrm{d} y- \frac{e}{\hbar}\iint\limits_{\mathbf{S}} \mathbf{B}\mathrm{d} \mathbf{A}~.
\label{Eq_Proj}
\end{equation}
The first integral is the projection of the electrostatic potential $V$ along the beam ($y$-)direction constricted by the local lamella thickness $t(x,z)=t_u-t_l$. The interaction constant $C_E$ is about 0.0065~(Vnm)$^{-1}$ at 300~kV. Being proportional to the magnetic flux of a magnetic induction $\mathbf{B}=\mu_0 \mathbf{H}_{\mathrm{d}}+\mu_0 \mathbf{M}$ through the surface $\mathbf{S}$ enclosed by the object- and the reference beam, the second integral quantifies the magnetic contribution to the phase.

Flipping the sample upside down\cite{Tonomura1986} for a second measurement yields $\varphi_\mathrm{flipped}$, which can be used to separate the electric $\varphi_\mathrm{el}$ and magnetic phase shift $\varphi_\mathrm{mag}$ as shown in Fig.~\ref{phase}(a-b):
\begin{eqnarray}
\varphi_\mathrm{el}&=&\frac{1}{2}\left(\varphi+\varphi_\mathrm{flipped}\right)\label{Eq_PhaseEl}\\
\varphi_\mathrm{mag}&=&\frac{1}{2}\left(\varphi-\varphi_\mathrm{flipped}\right)\label{Eq_PhaseMag}
\end{eqnarray}
As evident from Eq.~(\ref{Eq_Proj}), the electric phase contains the full information about the 3D sample geometry, which was further used to rebuild the structure for micromagnetic simulations. As another implication, the gradient of the magnetic phase returns purely the projected in-plane components of the magnetic induction:
\begin{equation}
\label{Eq_ProjMagTO}
\begin{split}
\negthickspace &\nabla\varphi_\mathrm{mag}(x,z)=\frac{e}{\hbar}\int\limits_{-\infty}^{+\infty}\negthickspace\mathbf{B}\negmedspace\times\negmedspace\mathrm{d} \mathbf{y}\\
&\negthickspace=\!\mu_0\frac{e}{\hbar}\!\!\left(\!\!\int\limits_{~t_l(x,z)}^{~t_u(x,z)}\!\negthickspace\negthickspace\negthickspace\negthickspace(\mathbf{M}\!+\!\mathbf{H}^\mathrm{int}_{\mathrm{d}})\negmedspace\times\negmedspace\mathrm{d}\mathbf{y}\!+\negmedspace\negthickspace\negthickspace\negthickspace\negthickspace\int\limits_{~t_u(x,z)}^{\infty}\!\negthickspace\negthickspace\negthickspace\negthickspace\mathbf{H}^\mathrm{ext}_{\mathrm{d}}\negmedspace\times\negmedspace\mathrm{d}\mathbf{y}\!+\negmedspace\negthickspace\negthickspace\negthickspace\negthickspace\int\limits_{-\infty}^{~t_l(x,z)}\!\negthickspace\negthickspace\negthickspace\negthickspace\mathbf{H}^\mathrm{ext}_{\mathrm{d}}\negmedspace\times\negmedspace\mathrm{d} \mathbf{y}\negthickspace\right)
\end{split}
\end{equation}
To obtain the internal demagnetizing field $\mathbf{H}^{\mathrm{int}}_{\mathrm{d}}$, a decomposition of $\mathbf{B}$ into the magnetization $\mathbf{M}$ and the demagnetizing field $\mathbf{H}_{\mathrm{d}}$ is necessary. A deeper technical description of the acquisition of a TEM hologram is provided in Refs.~[\onlinecite{Lehmann2002,Roeder2014}].

\begin{figure}[t]
\includegraphics[width=0.9\linewidth]{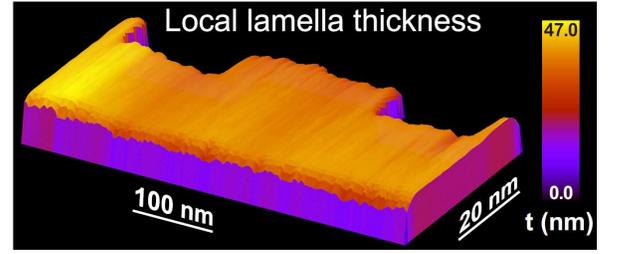}
\caption{\label{thickness} (Color online) Local lamella thickness $t$ in beam direction determined from the electric phase depicted in Fig.~\ref{phase}(a).}
\end{figure}

\vspace{3mm}
\textbf{(iii) Internal Demagnetizing Field Extraction}\quad 
After removing the phase contributions of the external strayfield (see Sec.~\ref{EH}), the vertically averaged distribution of the internal $H_{\mathrm{d},x}$ field was obtained by employing a numerical mask inside the magnetic region. In order to reduce the number of artifacts, areas of large phase noise were neglected. To achieve absolute field values in Tesla, the integrated magnetic phase was divided by the local lamella thickness (shown in Fig.~\ref{thickness}). Here, the field was averaged with the length of the vertical integration path and a Gaussian filter was applied in order to improve the signal-to-noise ratio of the extracted field distribution in Fig.~\ref{Hd}(d).

To reconstruct the $H^{\mathrm{int}}_{\mathrm{d},x}(x)$-distribution of an extended SMMC, a field scaling was necessary (see Sec.~\ref{EH}) due to two reasons. First, systematic deviations between the thickness-varied and Gaussian filtered simulation (solid orange line in Fig.~\ref{Hd}(d)) and the simulation of a perfectly flat lamella (dashed orange line in Fig.~\ref{Hd}(d)) were quantified and corrected. Second, the systematic differences of the internal field in a flat ($t_{\mathrm{avg}}=38.3$~nm thick) 2D structure compared to the field in a 3D magnonic crystal needed to be regarded.
%effects of the varying thickness and the were corrected by comparing the static simulation of a flat (quasi 2D) lamella with another one considering thickness variations (orange lines in Fig.~\ref{Hd}(d)).
%Therefore, initially the field-effects arising from the thickness variations were corrected employing the two magnetostatic simulations with and without considering the thickness variations (orange lines in Fig.~\ref{phase}(d)). 
Therefore, a scaling function was defined based on the static simulation of a flat quasi 2D lamella (dashed orange line in Fig.~\ref{Hd}(d)) and a 3D SMMC (dashed orange line in Fig.~\ref{Hd}(e)). Since the field values differ by more than one order of magnitude, the scaling was performed logarithmically:
\begin{equation}\label{Eq_scaling}
H^{\mathrm{3D}}_{\mathrm{d},x}(x) = \frac{H^{\mathrm{2D}}_{\mathrm{d},x}(x)}{\left|H^{\mathrm{2D}}_{\mathrm{d},x}(x)\right|}\cdot\left|H^{\mathrm{2D}}_{\mathrm{d},x}(x)\right|^{\left(\frac{\log{\left|H^{\mathrm{3D,sim}}_{\mathrm{d},x}(x)\right|}}{\log{\left|H^{\mathrm{2D,sim}}_{\mathrm{d},x}(x)\right|}}\right)}
\end{equation}
Here, $H^{\mathrm{3D}}_{\mathrm{d},x}(x)$ denotes the resulting 3D-corrected field measurement and $H^{\mathrm{2D}}_{\mathrm{d},x}(x)$ is the measured distribution of the thin (2D) lamella. The same field distributions obtained by simulations are labeled $H^{\mathrm{3D,sim}}_{\mathrm{d},x}(x)$ and $H^{\mathrm{2D,sim}}_{\mathrm{d},x}(x)$, respectively. Note that the index `int' was omitted in Eq.~(\ref{Eq_scaling}).

\vspace{3mm}
\textbf{(iv) Static Simulations}\quad 
For a thorough reconstruction of the lamella structure, static simulations were carried out. First, the average thickness $t_{\mathrm{avg}}$ of a flat lamella was varied until the magnetic phase inside the MC matched the mean phase obtained by measurement. With the help of that, the variations of the electric phase (Fig.~\ref{phase}(a)) inside the MC could be translated into local thickness variations with the result shown in Fig.~\ref{thickness}.
%The lateral cross-sectional shape depicted in Fig.~\ref{phase}(a) and the local lamella thickness shown in Fig.~\ref{thickness} were taken as a basis for the simulation geometry.
%From the electric phase image (Fig.~\ref{phase}(a)) the local lamella thickness profile shown in Fig.~\ref{thickness} was extracted. 
%depicted in Fig.~\ref{sim}. 
In order to consider tiny thickness variations in the static simulations, the saturation magnetization was scaled locally by $M^{'}_{\mathrm{S}}(x,y)=t(x,y)/t_{\mathrm{avg}}\cdot M_{\mathrm{S}}$ with a cell size of 2.438~nm\,$\cdot$\,2.125~nm$\,\cdot$\,2.410~nm for a high resolution. 
%This distribution of $M^{'}_{\mathrm{S}}(x,y)$ is given as black-white contrast in Fig.~\ref{sim}. 
Note that the thickness along the beam axis was fixed to the average value of $t_{\mathrm{avg}}=38.3$~nm. $M_{\mathrm{S}}=735$~kA/m, $D=23.6$~Tnm$^2$ and the $g$-factor $g=2.11$ were selected according to the material parameters of a permalloy reference film.\cite{Langer2016}

In order to compare a perfect (flat) 2D lamella with a 3D SMMC, the micromagnetic model above was modified omitting the local scaling of $M_{\mathrm{S}}$ with and without periodic boundary conditions in $y$-direction.

\vspace{3mm}
\textbf{(v) Dynamic Response Simulations}\quad 
The dynamic response simulations\cite{Vansteenkiste2014} were performed in two different ways. In order to obtain frequency-field dependencies (see Fig.~\ref{lamellafH}), pulsed\cite{McMichael2005} simulations were calculated. To simulate the angle-dependent spin-wave resonance (shown in Fig.~\ref{azimuth}), a continuous-wave approach\cite{Wagner2015} was chosen. As the latter does not require Fourier-transformations in frequency-space, such simulations could directly be carried out at $f=12$~GHz. 

Two different simulation geometries were selected: (i) a structural reconstruction of the shape of the magnonic crystal and (ii) an approach using the internal demagnetizing field $H^{\mathrm{int}}_{\mathrm{d},x}(x)$ only as an additive field in a 36~nm thin permalloy film. 

For the structural reconstruction of the magnonic crystal, the micromagnetic model according to the electrical phase image of the magnonic crystal (Fig.~\ref{phase}(a)) was applied. Minor changes of the simulation layout according to different \emph{average} values of $a_0=300$~nm and $w=166$~nm were regarded and, furthermore, the geometry was symmetrized. The modulation height was fixed to the value of $\Delta d=10$~nm with a continuous film of 26~nm thickness underneath. For an appropriate cross-sectional resolution, a cell size of $2.344~\mathrm{nm}\,\cdot\,4~\mathrm{nm}\,\cdot\,2~\mathrm{nm}$ was chosen with 128\,$\cdot$\,16\,$\cdot$\,18 cells in total. In order to realize a continuous elongation of the structure, the geometry was repeated 30 times in the $x$- and 100 times in the $y$-direction.

In the second approach, the internal demagnetizing field $H^{\mathrm{int}}_{\mathrm{d},x}(x)$ of the MC was added to an unmodulated continuous thin film. Due to the symmetry in $z$-direction, a larger cell-size of 12~nm was chosen with 3 cells in total along the $z$-axis. The cell size and the cell number along the $x$- and $y$-axis as well as the 2D repetitions were selected equivalently.
%\includegraphics[width=0.9\linewidth]{sim.eps}
%\caption{\label{sim} micromagnetic structure employed to model the lamella shown in Fig.~\ref{phase}(a-b). The varying thickness $t$ was implemented by a scaling of the local saturation magnetization %$M_{\mathrm{S}}$ according to the percentage given in black-white contrast.}
%\end{figure}
%
%\begin{figure}[tb]
%\includegraphics[width=0.9\linewidth]{Hd2.eps}
%\caption{\label{Hd2} (Color online) Colorplot of the $x$-component of the internal demagnetizing field obtained by micromagnetic simulations. The blue solid line shows the field distribution obtained by %vertical averaging.}
%\end{figure}
%
\bibliography{references}
\bibliographystyle{apsrev4-1}
\end{document}